\documentclass[twocolumn,showpacs,preprintnumbers,amsmath,amssymb,floatfix]{revtex4}

\usepackage[final]{graphicx}
\usepackage[caption = false]{subfig}
\usepackage{dcolumn}
\usepackage{bm}
\usepackage{array}
\usepackage{amsmath,amsfonts,amssymb}
\begin{document}

\title{Role of intersublattice exchange interaction on ultrafast longitudinal and transverse magnetization dynamics in Permalloy}

\date{\today}
\author{A. Maghraoui}
\author{F. Fras}
\author{M. Vomir}
\author{Y. Brelet}
\author{V. Halt\'e}
\author{J. Y. Bigot}
\author{M. Barthelemy}
\email[Corresponding author: ]{barthelemy@unistra.fr}
\affiliation{Universit\'e de Strasbourg, CNRS, Institut de Physique et Chimie des Mat\'eriaux de Strasbourg, UMR 7504, F-67000 Strasbourg, France}%

\begin{abstract}
 We report about element specific measurements of ultrafast demagnetization and magnetization precession damping in Permalloy (Py) thin films. Magnetization dynamics induced by optical pump at $1.5$eV is probed simultaneously at the $M_{2,3}$ edges of Ni and Fe with High order Harmonics for moderate demagnetization rates (less than 50\%). The role of the intersublattice exchange interaction on both longitudinal and transverse dynamics is analyzed with a Landau Lifshitz Bloch description of ferromagnetically coupled Fe and Ni sublattices. It is shown that the intersublattice exchange interaction governs the dissipation during demagnetization as well as precession damping of the magnetization vector.

\end{abstract}
\maketitle
\section{Introduction}
Ultrafast demagnetization of ferromagnets induced by femtosecond laser pulses 
\cite{Beaurepaire1996,Hohlfeld1997,Aeschlimann1997} promises novel applications in data storage and processing technologies. Since its discovery, several microscopic mechanisms such as the spin-orbit interaction \cite{Zhang2000,Bigot2009,Krieger2015}, Elliott-Yafet scattering induced spins-flips \cite{Koopmans2010}, non-thermal excitations \cite{Guidoni2002,Carva2013}, super-diffusive \cite{Battiato2010,Battiato2012} or ballistic spin-transport have been identified to play a key role and their relative weight can be element dependent \cite{Shokeen2017}. Depending on magnetic anisotropies, such transient modification of the effective magnetic field can trigger a coherent precession motion of the magnetization vector with a Gilbert damping \cite{Bigot2005} resulting from dissipation of energy to an external bath. Those longitudinal and transverse relaxation processes set a natural limit to optical manipulation of magnetization from femtosecond to nanosecond time scales. If one aims to study the  dynamics over such a large temporal scale, the Landau-Lifshitz Bloch (LLB) model \cite{Garanin1990} in which the effective field contains the essential microscopic mechanisms is well adapted. Among them, the exchange interaction appears to be critical, but several aspects remain to be explored. In particular, in multi-compound materials, the intersublattice exchange interaction plays a crucial role on the resulting global dynamics, acting as a spin momentum transfert between sublattices during the demagnetization \cite{Mentink2012}. Over the last decade, it has been investigated experimentally thanks to chemical selectivity of XUV resonant probe of core levels of transition metals (TM) and rare earths (RE). Time resolved Xray magnetic circular dichroism (XMCD) \cite{Stamm2007,Radu2009,Radu2011,Boeglin2010,Bergeard2014} and table-top high order harmonics (HH) probed time resolved magneto optical Kerr experiments (TMOKE)\cite{La-O-Vorakiat2009,Valencia2012,La-O-Vorakiat2012,Mathias2012,Gunther2014,Hofherr2018,Gang2018,Hofherr2020,Willems2020} have proven to offer a unique opportunity to study sublattice magnetization dynamics governed by dissipation and momentum transfert mechanisms in all optical magnetization switching in alloys \cite{Boeglin2010, Radu2011,Bergeard2014}. In particular, the demagnetization of each sublattice in a binary alloy can be either accelerated or decelerated compared to the pure element demagnetization \cite{La-O-Vorakiat2009,La-O-Vorakiat2012,Mathias2012}. This effect is dependent on the value of elemental magnetic momenta and on the ferro or antiferromagnetic nature of the exchange coupling \cite{Carley2012,Radu2015}. The case of Permalloy (Py) has attracted attention since various dynamical behaviors of sublattices magnetic momenta have been observed depending on the photon energy range of the probe. On one side, XMCD studies performed at $L_{2,3}$-edges have shown a faster demagnetization of Ni momenta compared to Fe \cite{Radu2015}. This observation is supported by the strong effective exchange coupling sustained by Fe momenta in Py so that Ni sublattice momenta are more submitted to thermal dissipation \cite{Hinzke2015}. On the other side, HH TMOKE measurements show that during the early demagnetization of ferromagnetic Py, the momenta of Fe starts to randomize before Ni momenta until a time of scale 10 fs after which both sublattice relax together due to  intersublattice exchange interaction (IEI) \cite{Mathias2012}. The origin of a stronger coupling of Fe spins to the electronic system compared to Ni remains unknown and deserves further exploration. In the present work, the magnetization dynamics of Fe and Ni sublattices of a $10$ nm Permalloy thin film is studied with chemical selectivity over a wide temporal range as a function of excitation density. A table top HH TMOKE configuration is used to measure both demagnetization and precession at the M edges of Fe and Ni. The role of strong intersublattice exchange interaction on longitudinal and transverse ultrafast magnetization dynamics is discussed for moderate demagnetization amplitudes.

\section{Experiment}
In our experiment, XUV sub $10$ fs pulses are produced by HH generation in a Ne-filled gas cell driven by  $795$ nm, $3$ mJ, $1$ kHz, $25$ fs laser pulses. The resulting XUV probe photons energies cover the 30 eV - 72 eV range and span the $M_{2,3}$-edges of Fe and Ni centered respectively at $66$ eV and $54$ eV.  Ultrafast demagnetization is induced by $795$ nm, $25$ fs pump with variable fluence in a $10$ nm thick Ni$_{80}$Fe$_{20}$ (Py) thin film with an in-plane anisotropy deposited on a crystalline Al$_2$O$_3$ substrate by ion beam sputtering.

\begin{figure}[htp]
\includegraphics[width= 8cm]{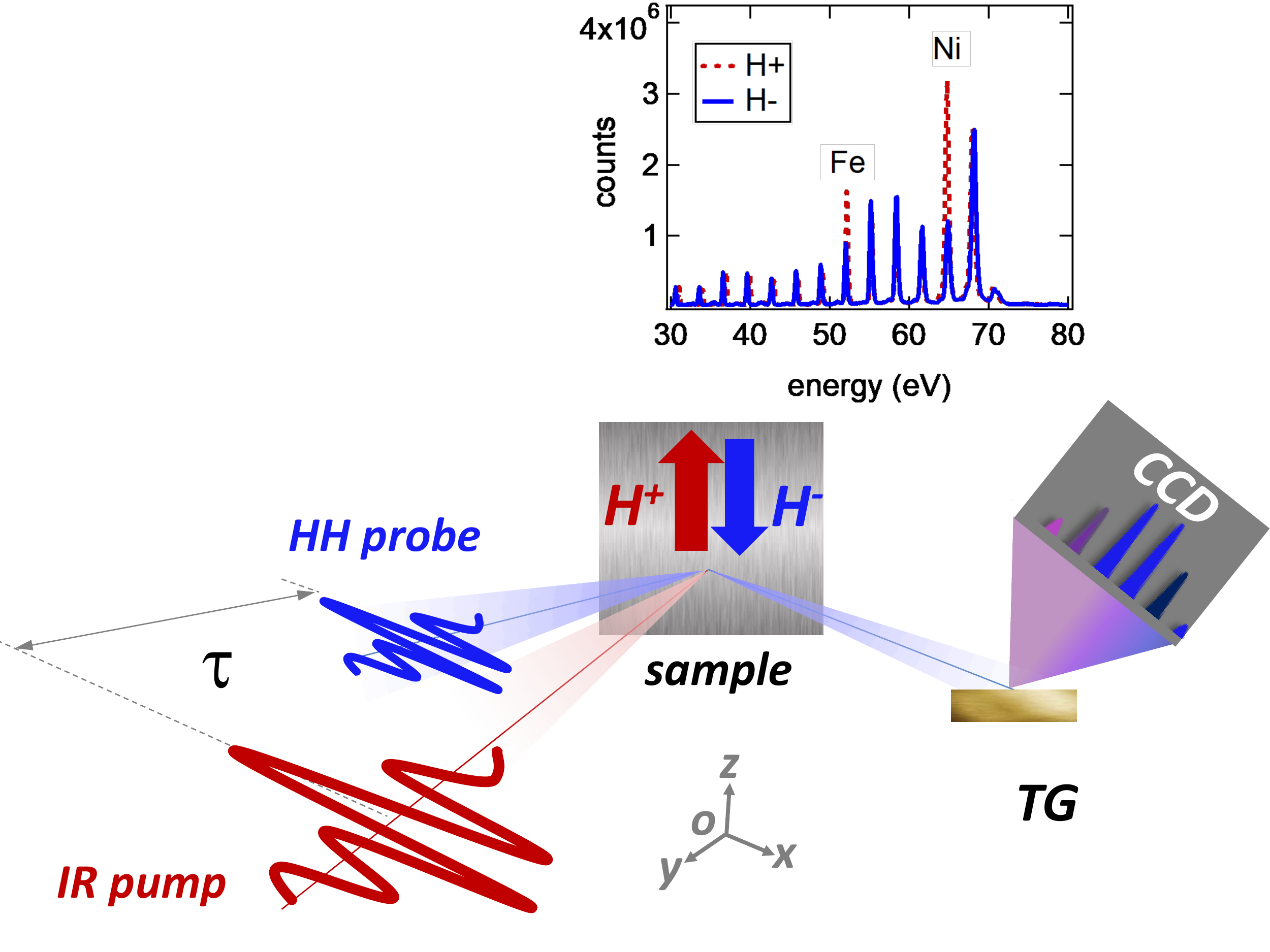}
\caption{Principle of XUV probe IR pump transverse HH TMOKE experimental configuration with static magnetic field H along $z$ axis. TG: toroidal grating. CCD: Charge Coupled Device camera. Inset: example of reflected spectra on Py for the antiparallel orientations  of the applied magnetic field +/-H (respectively blue solid line and red dotted line). }
\label{fig1}
\end{figure}

Figure \ref{fig1} illustrates the transverse time resolved magneto-optical Kerr configuration used in this work. An external static magnetic field ($H=450$ Oe) is applied on the sample along the transverse axis, $i.e.$ along z direction in figure \ref{fig1} and perpendicularly to the plane of incidence xOy of the p-polarized IR pump and VUV probe. The angle of incidence of the probe was set to  $45^{\circ}$ with respect to the sample normal in order to maximize the magnetic contrast obtained from spectrally resolved reflectivity measurements \cite{La-O-Vorakiat2009}. In the inset of figure \ref{fig1}, the reflected XUV probe spectra $I_{H^+}^{stat}$ and $I_{H^-}^{stat}$ is shown for two antiparallel orientations of the transverse magnetic field $H$. The maximum intensity difference between the two reflected spectra is seen at the harmonics $h_{45}$(centered at $66$ eV) and $h_{35}$(centered at $54$ eV) corresponding  to the M-edges of Ni and Fe respectively. Both spectra are further measured as a function of pump probe delay by varying the optical path of the pump with a mechanical delay line.

\section{Ultrafast demagnetization in Permalloy probed at M-edges of Ni and Fe}
We first consider the short time scale corresponding to demagnetization process in Permalloy. The elemental magnetization dynamics of Ni and Fe elements $\text{m}(q,\tau)$ measured as a function of the pump-probe delay $\tau$ is then integrated over each resonant $q^{th}$ harmonic:
\begin{equation}
\frac{\Delta \text{m}}{\text{m}}(q,\tau)=\frac{I_{H^+}^{\text{dyn}}(q,\tau)-I_{H^-}^{\text{dyn}}(q,\tau)}{I_{H^+}^{\text{stat}}(q)-I_{H^-}^{\text{stat}}(q)}
\end{equation}
for $q=45$ and $q=35$ with $I_{H^\pm}^{\text{dyn}}=I_{H^\pm}^{\text{with}}-I_{H^\pm}^{\text{stat}}$ and $I_{H^\pm}^{\text{with}}$ being the intensity of signal with pump and $I_{H^\pm}^{\text{stat}}$ without pump.
\begin{figure}[htp]
     \includegraphics[width = 8cm]{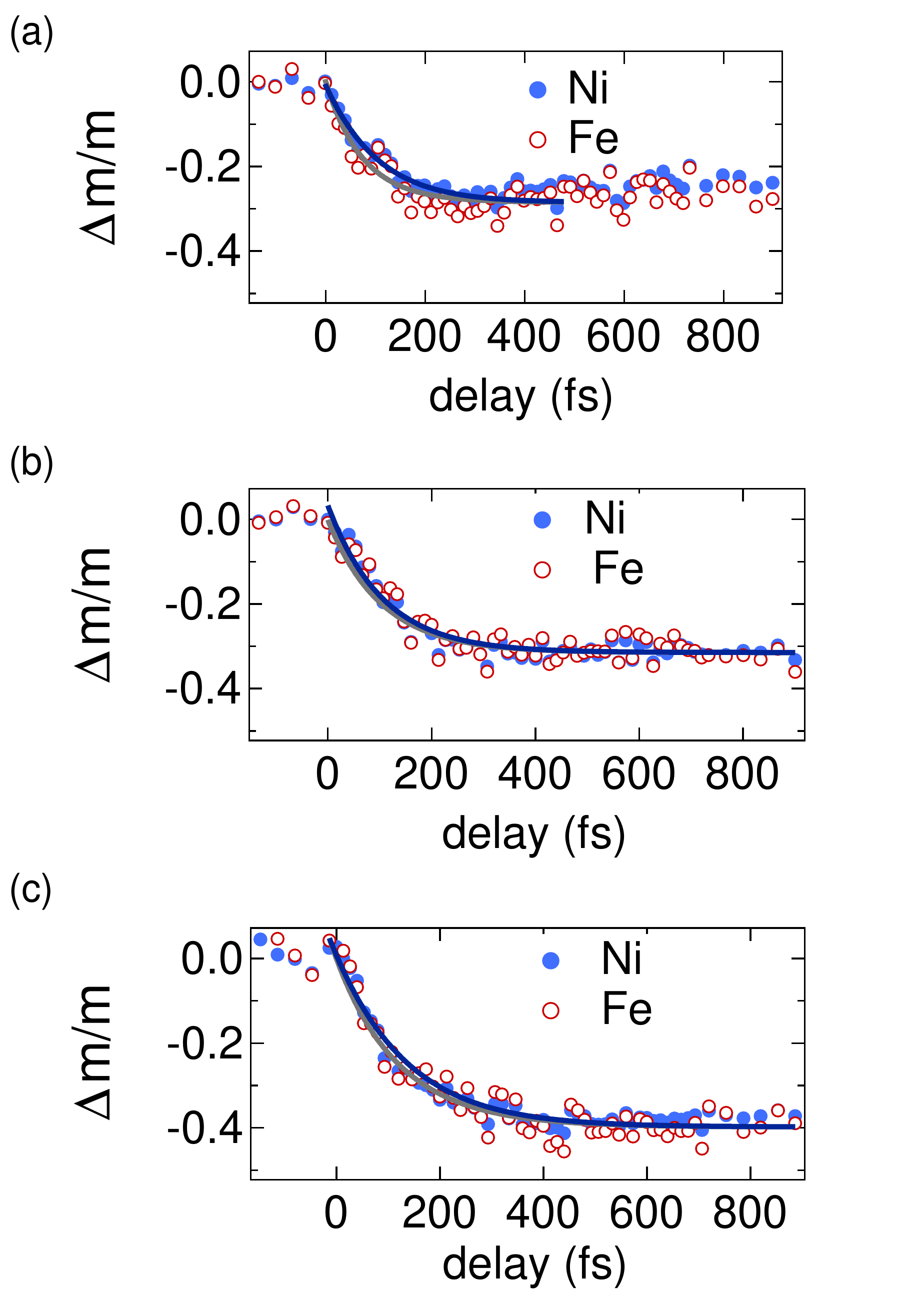}
           \caption{Demagnetization dynamics in permalloy probed at the Ni (full blue dots) and Fe (empty red dots) M-edges and fits (grey lines: Fe, blue lines: Ni) for incident pump fluences of a) $2.5$ mJ/cm$^2$.  b) $3.8$ mJ/cm$^2$. c) $4.7$ mJ/cm$^2$.}
\label{fig2}
\end{figure}
Figure \ref{fig2} shows demagnetization $\frac{\Delta \text{m}}{\text{m}}(q,\tau)$ in Py at the $M_{2,3}$ edges of Fe ($\frac{\Delta \text{m}^{Fe}}{\text{m}^{Fe}}$) and Ni  ($\frac{\Delta \text{m}^{Ni}}{\text{m}^{Ni}}$) integrated over harmonics $h_{35}$ and $h_{43}$ respectively for three increasing pump fluences. Ni and Fe sublattices appear to demagnetize simultaneously. The demagnetization amplitude of both sublattices increases from 25 $\%$ to 40 $\%$. Contrary to reference \cite{Mathias2012}, no reproducible delay between the two sublattices demagnetizations is observed with our pump duration of 25 fs. A possible explanation could be  a slight variation of intersublattice exchange interaction (sample dependent due to change of crystallinity or grating vs alloy) that may induce a change of the temporal shift value. Moreover the different conditions of HH generation could lead to a different time duration of our probe resulting to a lower temporal resolution, or a delay between h35 and h45 due to a possible chirp.

The framework of analysis and data fitting is described in the following, where details about linearized LLB are presented. This approach is based on the knowledge of the laser induced temperature of the system. In order to define such laser induced temperature, let us first explore the demagnetization amplitude behavior with incident pump fluence.

\begin{figure}
\includegraphics[width = 8cm]{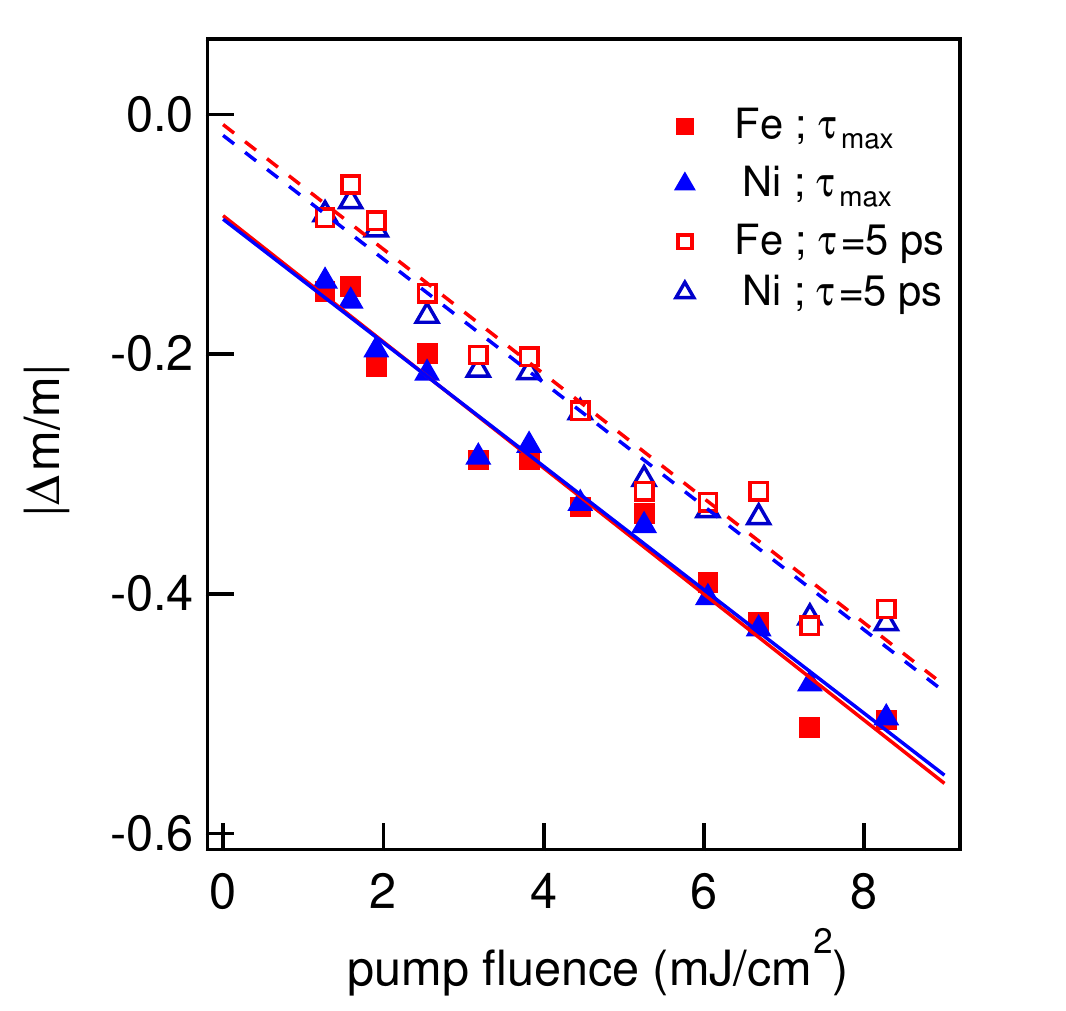}
\caption{Demagnetization amplitudes measured at M-edges of Ni and Fe in Py as a function of incident pump fluence. For maximum demagnetization (full symbols) and at a $5$ ps pump probe delay (empty symbols).}
\label{fig3}
\end{figure}

The demagnetization amplitude of Ni and Fe in permalloy with respect to the pump fluence is plotted in figure \ref{fig3}. For each point, two delays have been chosen. The first delay corresponds to maximal demagnetization level (the corresponding time delay range is 300 fs - 500 fs from low to large pump fluences), the second delay is fixed at 5 ps and corresponds to a thermal quasi-equilibrium between electrons and lattice baths (remagnetization). Within the fluence range of this experiment the demagnetization amplitude is always found identical for both sublattices.
Moreover, up to 40$\%$ demagnetization given by the damage threshold of our sample, a linear increase of demagnetization amplitude with the pump fluence is observed, with the same slope for both delays.
Such linear behavior can be attributed to a coupling with a thermalized bath\cite{Hohlfeld1997}, characterized by an energy of $k_B T$, heated by the laser pulse. 
The vertical offset between the two curves arises due to the magnetization recovery. In LLB approach, it can be described as a temperature step between maximal bath temperature (demagnetization) and cooling down of the bath due to coupling with lattice (remagnetization).
One can notice that the linear fits labelled “maximum” does not cross the zero variation at zero fluence. At a delay of 5 ps a slight deviation from zero is also observed. This range of fluence is either difficult to access experimentally or usually not considered. A hypothesis to explain such a behavior is a change of the regime of interaction, at the origin of the demagnetization or remagnetization processes, with pump fluence. This could lead to a different power dependent law in the range of very low fluences, but this aspect goes beyond the scope of the present work.
After thermalization of electrons, the temperature dependent magnetization can be approximated with a molecular field model as :
\begin{equation}
\mathbf{m}^{\epsilon}(T)=(1- \frac{T}{T_C})^{\kappa^\epsilon}
\label{eq3}
\end{equation}
where $T_C$ is the Curie temperature, $\kappa^{\epsilon}$ is the critical exponent. In the following, the maximum amplitude of demagnetization in our experiments is used to evaluate the laser induced spin temperature $T$ that is related to the amplitude of fluctuations to which the spins are submitted \cite{Atxitia2012} in the 300-500 fs temporal range. As in ref \cite{Hinzke2015}, in our approach at short time scale, only the demagnetization process is considered, justified by a slower rate of the re-magnetization process. The magnetic system can be considered initially in thermal equilibrium at a temperature of $T_i$ = 300K, then for t=0 the bath temperature is instantaneously changed to $T_f$. Thus, the magnetization of the two sublattices will evolve towards a new thermal equilibrium value given by $T_f$. Therefore, in this approach, the equivalent spin temperature is defined at the maximum of demagnetization. By taking a Curie Weiss law ($\kappa=1/2$) and $T_C$ = 850 K in Ni$_{80}$Fe$_{20}$, the temperature range can be calibrated. In figure \ref{fig3}, the evaluated spin temperature $T$ ranges from 300 K to 600 K, and the maximum demagnetization amplitudes of figure \ref{fig2} correspond to $T/T_C$ = 0.35, 0.43 and 0.52.

We now analyze our experimental data in the frame of the linearized LLB model. It considers an ensemble of rigid spins submitted to exchange interaction and coupled to a thermal bath
corresponding to either charges or phonons at the origin of dissipation \cite{Atxitia2011,Atxitia2012,Schellekens2013,Wienholdt2013, Nieves2014}. It gives a consistent approach that encompasses a broad temporal scale from femtoseconds to nanoseconds during which spin flips as well as the magnetization precession take place. By isolating the longitudinal contribution to
magnetization dynamics at short time scales, it can be used to simulate ultrafast demagnetization in TM-RE compounds. This method was first applied to ferrimagnets known for their
high potential for all optical switching \cite{Atxitia2012}, and more recently to better understand the role of intersublattice exchange interaction (IEI) in ferromagnetic TM alloys
\cite{Hinzke2015}. In particular, this model allows to decipher quantitatively the role played by both the IEI and intrinsic dissipation of each sublattices magnetization on the
observed dynamics. Fe and Ni momenta dynamics in permalloy are described with the following first order coupled rates equation:
\begin{equation}
\left(
\begin{array}{c}
\dot{\mathbf{m}}^{\text{Fe}} \\
\dot{\mathbf{m}}^{\text{Ni}}
\end{array}
\right)
=\cal A_{\|}
\left(
\begin{array}{c}
\mathbf{m}^{\text{Fe}} \\
\mathbf{m}^{\text{Ni}}
\end{array}
\right)
= 
\left(
\begin{array}{cc}
-\Gamma_{\text{FeFe}}& \Gamma_{\text{FeNi}}\\
\Gamma_{\text{NiFe}}& -\Gamma_{\text{NiNi}}
\end{array}
\right)
\left(
\begin{array}{c}
\mathbf{m}^{Fe} \\
\mathbf{m}^{Ni}
\end{array}
\right)
\label{eq7}
\end{equation}
where the matrix $\cal A_{\|}$ drives the dynamics. Its elements can be written as a function of micromagnetic parameters\cite{Hinzke2015}:
\begin{align}
\Gamma_{\text{FeFe}}=&\frac{1}{\tau^{\text{Fe}}_{\text{intra}}} + \frac{\chi_{\|}^{\text{Ni}}}{\chi_{\|}^{\text{Fe}}}\frac{1}{\tau^{\text{FeNi}}_{\text{exch}}}
\label{eq15a}\\
\Gamma_{\text{FeNi}}=&\frac{1}{\tau^{\text{FeNi}}_{\text{exch}}}
\label{eq15b}\\
\Gamma_{\text{NiFe}}=&\frac{1}{\tau^{\text{NiFe}}_{\text{exch}}}
\label{eq15c}\\
\Gamma_{\text{NiNi}}=&\frac{1}{\tau^{\text{Ni}}_{\text{intra}}} + \frac{\chi_{\|}^{\text{Fe}}}{\chi_{\|}^{\text{Ni}}}\frac{1}{\tau^{\text{NiFe}}_{\text{exch}}}
\label{eq15d}
\end{align}

In this basis, the elements of $\cal A_{\|}$ contain two contributions to longitudinal damping. The first one corresponds to intrasublattice demagnetization $\tau_{\text{intra}}$:
\begin{equation}
\frac{1}{\tau_{\text{intra}}^{\text{Fe,Ni}}}=\frac{\gamma^{\text{Fe,Ni}}\alpha^{\text{Fe,Ni}}_{\|}(T)}{\chi_{\|}^{\text{Fe,Ni}}(T)}
\label{eq9}
\end{equation}

where $\alpha^{\text{Fe,Ni}}_{\|}$ is longitudinal damping and $\chi_{\|}^{\text{Fe,Ni}}$ is the effective magnetic susceptibility, both being element and temperature dependent quantities. $\gamma^{\text{Fe,Ni}}$ corresponds to the gyromagnetic ratio of each element.
The second one is the intersublattice exchange mediated demagnetization:

\begin{equation}
\frac{1}{\tau_{\text{exch}}^{\text{FeNi,NiFe}}}=\frac{\gamma^{\text{Fe,Ni}}\alpha^{\text{Fe,Ni}}_{\|}(T)J_{\text{FeNi,NiFe}}}{\mu^{\text{Fe,Ni}}}
\label{eq10}
\end{equation}
 where $J_{\text{FeNi}}=J_{\text{NiFe}}$ is the IEI constant and $\mu_{\text{Fe,Ni}}$ is the atomic magnetic momentum. 
 
  Finally, diagonal terms of $\cal A_{\|}$, $-\Gamma_{FeFe}$ and $-\Gamma_{NiNi}$,  correspond to a dissipation of magnetic momenta in each sublattice and via IEI with the second sublattice (eq. \ref{eq15a}). Non  diagonal terms $\Gamma_{NiFe}$ and $\Gamma_{FeNi}$ lead to an exchange of momentum between sublattices. It should be underlined that, in this approach, the overall flow of momentum, mediated by the IEI, is element dependent and weighted by the magnetic susceptibilities ratio.
  
The differential system (\ref{eq7}) can be easily solved in the eigen basis after diagonalization of $\cal A_{\|}$:
\begin{equation}
\cal A_{\|}=
\left(
\begin{array}{cc}
\Gamma^+ & 0\\
0& \Gamma^-
\end{array}
\right)
\end{equation}
where the two eigen values $\Gamma^{\pm}=1/\tau^{\pm}$ can be written as:
\begin{multline}
\Gamma^{\pm}=\frac{1}{2}(\Gamma_{\text{FeFe}}+\Gamma_{\text{NiNi}}\\
\pm \sqrt{(\Gamma_{\text{FeFe}}-\Gamma_{\text{NiNi}})^2+4\Gamma_{\text{FeNi}}\Gamma_{\text{NiFe}}})
\label{eq14}
 \end{multline}
Finally the measured sublattices magnetization dynamics can be expressed as a linear combination of the differential system solutions:
\begin{align}
\frac{\Delta \mathbf{m}^{\text{Fe}}}{\text{m}^{Fe}}= A^{\text{Fe}}\exp (-\frac{t}{\tau^+})+B^{\text{Fe}}\exp (-\frac{t}{\tau^-}),
\label{eq14a}\\
\frac{\Delta \mathbf{m}^{\text{Ni}}}{\text{m}^{Ni}}= A^{\text{Ni}}\exp (-\frac{t}{\tau^+})+B^{\text{Ni}}\exp (-\frac{t}{\tau^-}),
\label{eq14b}
\end{align}
where the coefficients $A^{Fe}$,$B^{Fe}$,$A^{Ni}$,$B^{Ni}$ depend on the eigen vector components $x^{\pm}=\Gamma_{\text{FeNi}}/(\Gamma_{\text{FeFe}}-1/\tau^{\pm})$ as follows:
\begin{eqnarray}
A^{Fe}=&\Delta m^{Fe}_0\frac{x^+}{(x^--x^+)}(\frac{\Delta m^{Ni}_0}{\Delta m^{Fe}_0}x^- -1)\\
B^{Fe}=&\Delta m^{Fe}_0\frac{x^+}{(x^--x^+)}(1- \frac{\Delta m^{Ni}_0}{\Delta m^{Fe}_0}x^+)\\
A^{Ni}=&\Delta m^{Fe}_0\frac{1}{(x^--x^+)}(\frac{\Delta m^{Ni}_0}{\Delta m^{Fe}_0}x^- -1)\\
B^{Ni}=&\Delta m^{Fe}_0\frac{1}{(x^--x^+)}(1- \frac{\Delta m^{Ni}_0}{\Delta m^{Fe}_0}x^+)
\label{eq13}
\end{eqnarray}

with $\Delta m^{Fe}_0$ and $\Delta m^{Ni}_0$ corresponding to the maximum amplitude of demagnetization.
It is important to notice that $\tau^{\pm}$ only corresponds to $\tau^{Fe,Ni}$ in the very low temperature range, when $1/\tau^{\text{NiFe}}_{\text{exch}}$ and $1/\tau^{\text{FeNi}}_{\text{exch}}$ are negligible. In the range of temperatures explored in our experiment, due to IEI, the dynamics of each sublattice is a clear bi-exponential decay as shown in eq.\ref{eq14a} and \ref{eq14b} where the demagnetization time is a composition of $\tau^-$ and $\tau^+$.
Having in hands the $T/T_C$ values equivalent to pump fluences from amplitudes of demagnetization of our experiment, we can analyze our results in the frame of the linearized LLB model. As shown earlier, the pump fluence range used in our experiments corresponds to an intermediate temperature range $0.35 < T/T_C < 0.52$.

\begin{figure}[htp]
\includegraphics[width= 9cm]{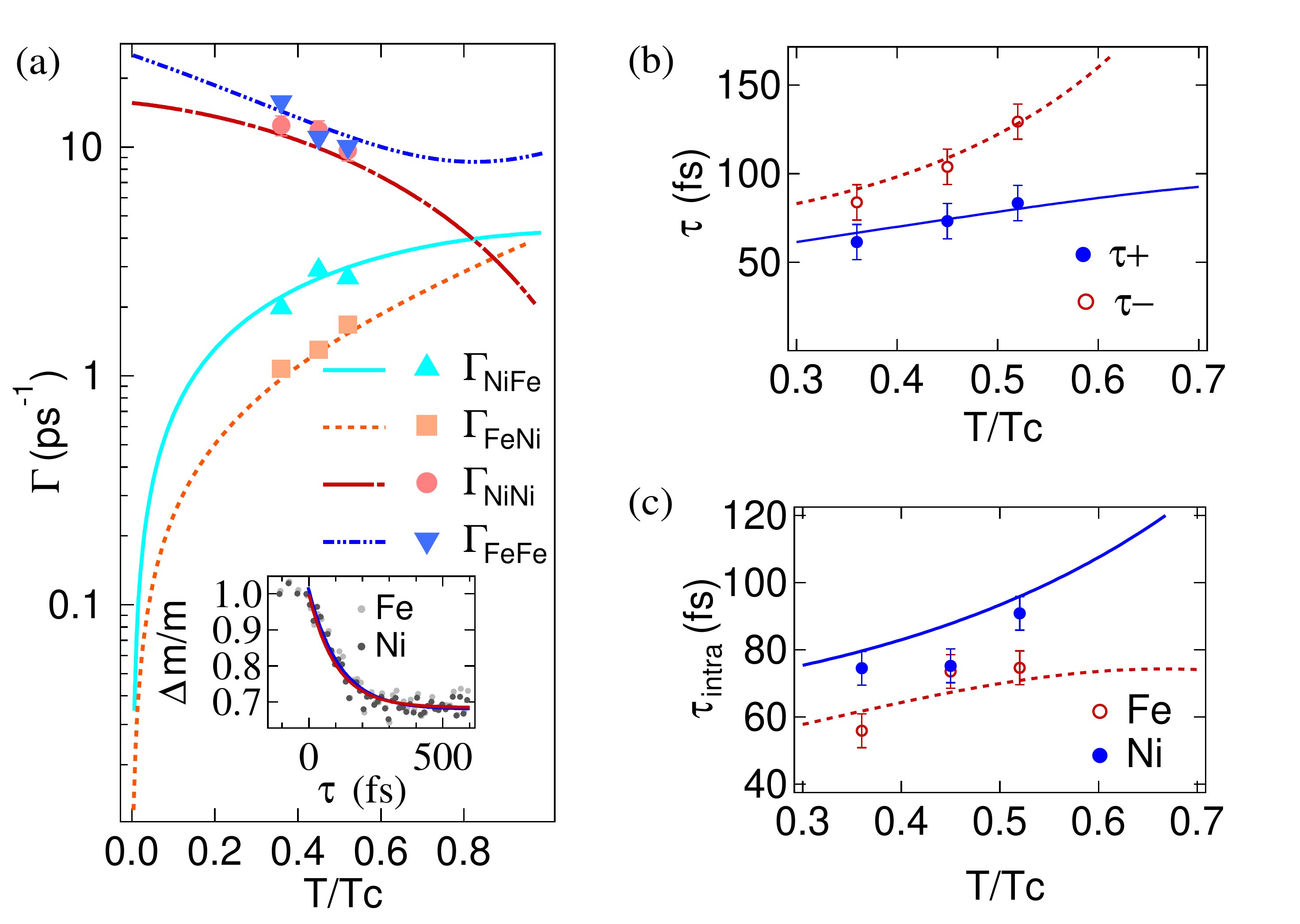}
\caption{Comparison of theoretical data from \cite{Hinzke2015} (lines) and experimental values (dots) of a) LLB matrix components for pump energy densities of $T/T_C=$ 0.35, 0.42 and 0.52. Experiment (markers) and theory (lines). Inset: example of M-edge demagnetization global fitting (lines) giving  rise to LLB matrix element for 3.8 mJ/cm$^{-2}$ pump fluence. b) characteristic times of M edges magnetization dynamics $\tau^+$ (empty red dots), $\tau^-$ (full blue dots) and c) retrieved $\tau^{Ni}_{\text{intra}}$ (full blue dots), $\tau^{Fe}_{\text{intra}}$ (empty red dots), from experiment and comparison with theoretical values as a function of temperature.}
\label{fig4}
\end{figure}
Figure \ref{fig4} shows a comparison between theoretical values obtained from multiscale LLB approach in \cite{Hinzke2015} and our experimental ones of (a) the subsequent components of the LLB matrix $\cal A_{\|}$ and (b) $\tau^+=1/\Gamma^+$, $\tau^-=1/\Gamma^-$, which are the characteristic times of the bi-exponential solutions of differential equations that govern the two sublattices dynamics.
For each pump fluence and corresponding laser induced temperature, a global fitting procedure is used simultaneously for both Fe and Ni M-edges demagnetization measurements to retrieve the $\cal A_{\|}$ matrix elements plotted as symbols in figure \ref{fig4}(a). From those values, $A^{\epsilon},B^{\epsilon}$ and $\tau^{\pm}$ can be calculated and injected in equations (\ref{eq14a}) and (\ref{eq14b}). An example of the corresponding $\Delta \mathbf{m}^{\text{Fe}}/\text{m}^{Fe}$ and $\Delta \mathbf{m}^{\text{Ni}}/\text{m}^{Ni}$ curves are plotted as lines in the inset of figure \ref{fig4}(a). It reproduces very well the observed demagnetization dynamics in Py measured at M-edges of $Fe$ and $Ni$ sublattices.
Moreover, a very good agreement is obtained between the theoretically predicted values of the $\cal A_{\|}$ elements and those extracted from our experiments fits (shown as lines in figure \ref{fig2}). The non diagonal elements of $\cal A_{\|}$ retrieved from experiments confirm that IEI mediated dissipation is much stronger for Ni sublattice compared to Fe ($\Gamma_{\text{NiFe}}>\Gamma_{\text{FeNi}}$). Indeed, the IEI induced modifications of both sublattice dynamics are not the same, due to element dependence of $\Gamma_{FeNi,NiFe}$ (via element dependence of $\alpha_{\|}^{Fe,Ni}$, $ \mu^{Fe,Ni}$ and $ \gamma^{Fe,Ni}$).
In order to discuss the intrasublattice dissipation (without the contribution of IEI), we extract the intrasublattice demagnetization time from our measurements, as shown in figure \ref{fig4}(c). It is deduced from the experimentally retrieved values of $\cal A_{\|}$ matrix elements and from equations \ref{eq15a} and \ref{eq15d} by taking a constant ratio of magnetic susceptibilities  $\chi_{\|}^{Fe}/\chi_{\|}^{Ni}=2$ (valid in the intermediate temperature range i.e $T/T_C<$0.5 \cite{Hinzke2015}). Up to $T/T_C$ = 0.52, $\tau^{Ni}_{\text{intra}}-\tau^{Fe}_{\text{intra}}<15$ fs. Without IEI (figure \ref{fig4}(c)), Fe sublattice undergoes a stronger dissipation. This disparity between intra sublattice dissipations is compensated by strong IEI leading to a common dynamics of both sublattices (figure \ref{fig4}(b)).
The above approach has the advantage to explain the observed dynamics strongly influenced by IEI. It shows that IEI mediated dissipation doesn't have necessarily the same weight on each sublattice. Moreover, the rate of intrasublattice dissipation mediated by IEI is related to sublattices magnetic susceptibilities ratio, that is almost constant for moderate laser induced temperature and diverges close to $T_C$. Let us discuss this substantial difference with the definition of a unique sub $10$ fs exchange time observed in previous work. 
Indeed, in the pioneer study proposed by Mathias et al \cite{Mathias2012}, an exchange interaction time is introduced as a constant parameter that couples the two subsystems magnetization dynamics. The key differences are based on the following points. In ref \cite{Mathias2012}, the process of IEI is considered in a conservative manner with equal rates of magnetic momentum transfer between the two sublattices. The exchange interaction times are equal for both sublattices and independent of fluence. Finally the data analysis, performed in this framework, imposes a strong difference between $\tau_{Fe}$ and $\tau_{Ni}$ (See supplementary informations of \cite{Mathias2012} ). The LLB based approach is fundamentally different since it considers the effect of IEI as related to the conservative transfer of momentum between sublattices, but also to dissipation (eq. \ref{eq15a} - \ref{eq15d}). Secondly, with LLB approach, these contributions are both found element and temperature dependent (eq. \ref{eq10}) due to longitudinal damping and magnetic susceptibilities. The outcome analysis allows retrieving the intrasublattice demagnetization times: between 80 and 100 fs for both Fe and Ni, which is consistent with earlier observations \cite{Koopmans2010,Walowski2008}.

\section{Ultrafast magnetization precession and damping in Permalloy probed at M-edges of Ni and Fe}

We have shown the influence of intersublattice exchange interaction on longitudinal magnetization dynamics occuring on the hundreds of femtoseconds time scale. We now address the question of how IEI does affect the damping of transverse motion of magnetization vector, ie precessional motion over hundreds of picoseconds. In the following, the IR pump XUV probed TMOKE experiments are performed on a 500 ps temporal range with a tilt of the external magnetic field axis with a $10^{\circ}$ angle with respect to the sample plane. This configuration allows for transverse projection measurement of magnetization precession, simultaneously at M$_{2,3}$ edges of Fe and Ni. Both Kerr rotation signals  $\Delta\theta_K/\theta_K$ integrated over $h35$ (Fe) and $h43$ (Ni) have been fitted using the fitting function: $A^{\text{Ni,Fe}}\text{sin}(2\pi/T_{\text{pr}}^{\text{Ni,Fe}}t+\phi^{\text{Ni,Fe}})\text{exp}(-t/T_{\text{d}}^{\text{Ni,Fe}})+B^{\text{Ni,Fe}}$, where $A^{\text{Ni,Fe}}$, $T_{pr}^{\text{Ni,Fe}}$,$\phi^{\text{Ni,Fe}}$, $T_{\text{d}}^{\text{Ni,Fe}}$  are respectively the precession amplitudes, periods, phases and damping times of each sublattice. $B^{\text{Ni,Fe}}$ is an offset that corresponds to long time delay magnetization recovery compared to the temporal window of our measurements.

\begin{figure}[htp]
\includegraphics[width= 7cm]{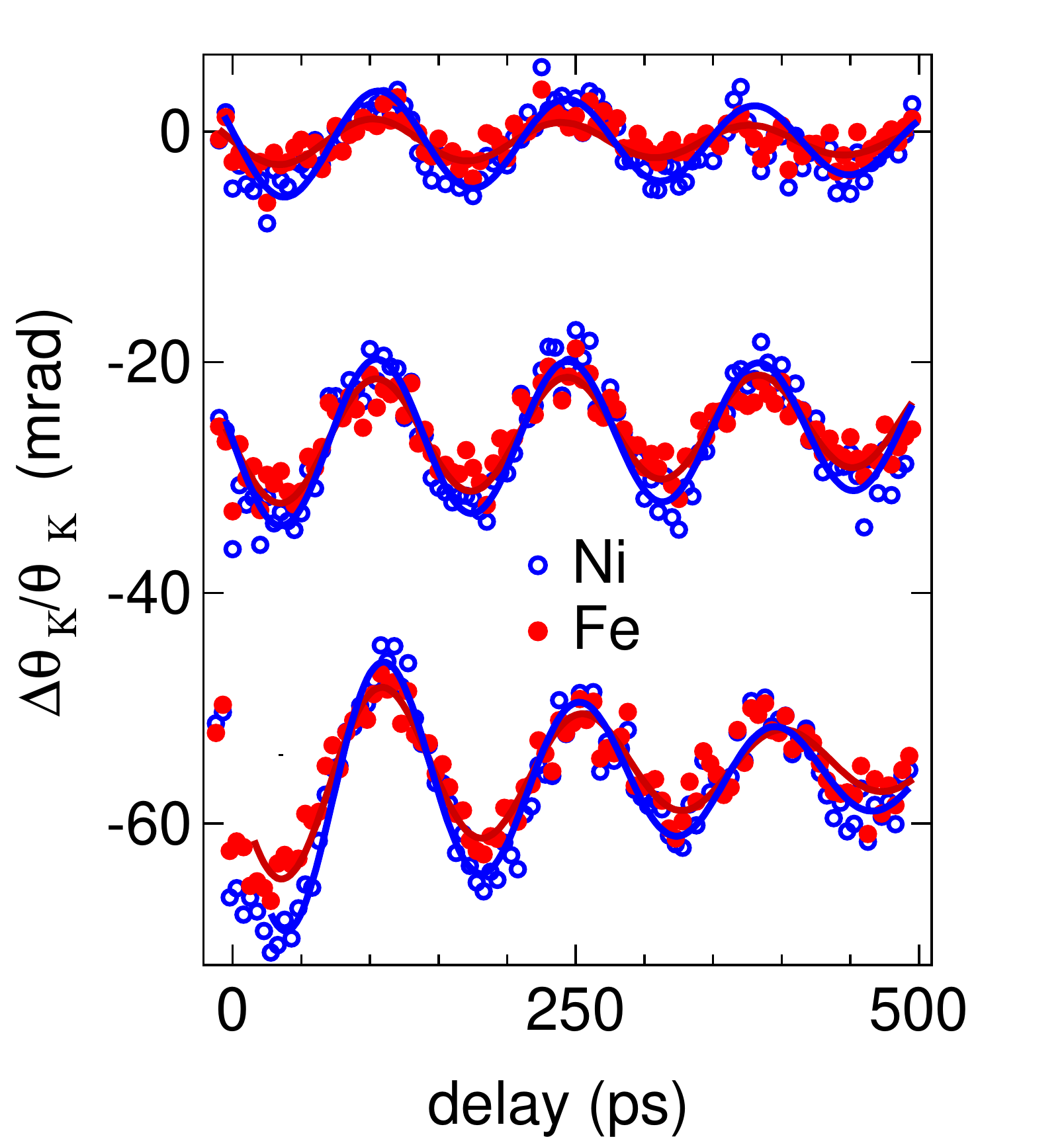}
\caption{Precession measurements in permalloy probed with High order Harmonics. Fe M-edge (full red dots) and Ni M-edge (empty blue dots) magnetization dynamics of precession in Permalloy as a function of incident pump laser fluence increased from top to bottom: $1.5$ mJ/cm$^2$ ; $2.5$ mJ/cm$^2$; $4$ mJ/cm$^2$.}
\label{fig5}
\end{figure}

As seen in figure \ref{fig5}, the Ni and Fe momenta precession are measured selectively in Ni$_{80}$Fe$_{20}$ for three incident pump fluences corresponding to initial $25\%$ to $35\%$ of laser induced demagnetization. The precession motion at Ni edge has a slightly higher amplitude. While the precession signals are increased in amplitude with pump fluence, Fe and Ni momenta still precess in phase.  Within the range of pump fluences used, the precession period stays quasi constant $T_{\text{pr}}^{\text{Ni,Fe}}\sim150$ ps. The damping time remains identical for both sublattices. It is found to be of about 2 ns for the intermediate fluence and is decreased down to 300 ps at the highest fluence. For the lowest fluence the damping time is difficult to extract within our  temporal window and lower signal to noise ratio. It is is estimated higher than 2 ns. Such increase of damping can be explained by the increasing of phonon mediated spin flip rate  with increasing fluence. Indeed, the generated phonon density increases with pump fluence. Phonons cause crystal field fluctuations that translates to the magneto-crystalline anisotropy and leads to random torques on the spins. 
We now analyze the Gilbert damping in the frame of the Landau Lifshitz Gilbert (LLG) equation. In the time scale of the transverse damping (three orders of magnitude longer compared to the short time scale of ultrafast demagnetization), the precession motion of two coupled sublattices $\epsilon$ and $\delta$ magnetization can be written for $\textbf{m}_{\epsilon, \delta}$ ($m^{\delta}_s$ being the saturation magnetization of the second sublattice $\delta$) as \cite{Schlickeiser2012}:
\begin{multline}
\frac{\dot{\textbf{m}}^{\epsilon}}{\gamma^{\epsilon}}=-\left(\textbf{m}^{\epsilon}\times \textbf{H}'_{\epsilon}\right)-\alpha^{\epsilon}\left[\textbf{m}^{\epsilon} \times \left(\textbf{m}^{\epsilon} \times \textbf{H}'_{\epsilon}\right)\right] \\
+ A_{\epsilon\delta} m^{\delta}_s \left\{ \left(  \textbf{m}^{\epsilon} \times \textbf{m}^{\delta} \right)+\alpha^{\epsilon}A_{\epsilon\delta} m^{\delta}_s\left[\textbf{m}^{\epsilon}\times \left(\textbf{m}^{\epsilon} \times \textbf{m}^{\delta}\right) \right]\right\}
\label{eq11}
\end{multline}

The first line of equation (\ref{eq11}) corresponds to precession of magnetization and damping related to effective field $\textbf{H}'_{\epsilon}=\textbf{H}_0+\textbf{H}_{anis}$, where $\gamma^{\epsilon}$ is the gyromagnetic ratio, $\alpha^{\epsilon}$ the Gilbert damping,$\textbf{H}_0$ and $\textbf{H}_{anis}$ are the applied and anisotropy fields. The second line corresponds to the coupling of precession motion and damping via IEI $J_{\epsilon \delta}$ between sublattices $\epsilon$ and $\delta$. It corresponds to a contribution to the effective field $\textbf{H}''_{\epsilon}=-A_{\epsilon\delta}m_s^{\delta} \textbf{n}^\delta$ of the second sublattice $\delta$ on the first sublattice $\epsilon$. The exchange stiffness parameter is defined by $A_{\epsilon\delta} = J_{\epsilon \delta}/\mu_{\delta}\mu_{\epsilon}$. 

Note that when $A_{\epsilon\delta}$ is lower than other elemental effective field contributions, the two sublattices precess independently. In our case, when $A_{\epsilon\delta}$ dominates, the resulting motion corresponds to a single coupled precession motion. The corresponding Gilbert damping can be evaluated from the damping time $T_{\text{d}}^{\text{Py}}$ as a single value for each fluence. Considering a circular precession motion with small angles, one has:
\begin{equation}
\alpha^{\text{Py}}=1/(T_{\text{d}}^{\text{Py}}\omega^{\text{Py}})
\label{eq12a}
\end{equation}
with $\omega^{\text{Py}}=2\pi/T_{pr}$ being the precession pulsation.
 From our measurements in Py at M-edges of Ni and Fe (figure \ref{fig5}) and by taking m$_{\text{s}}^{\text{Py}}$ = 8.4 10$^6$A m$^{-1}$, one has:  $\alpha^{\text{Ni,Py}}\sim\alpha^{\text{Fe,Py}}\leq 0.012$ for the two first fluences and $\alpha^{\text{Ni,Py}}\sim\alpha^{\text{Fe,Py}}=0.079$ for maximal fluence.
 Moreover, in a strongly exchange coupled alloy, its Gilbert damping can be estimated as an effective damping from pure elements parameters \cite{Gurevich1996}:
\begin{equation}
\alpha^{\text{Py}}_{\text{eff}}=\frac{m^{\text{Fe}}_s\gamma^{\text{Ni}}\alpha^{\text{Fe}}+m^{\text{Ni}}_s
\gamma^{\text{Fe}}\alpha^{\text{Ni}}}{m^{\text{Fe}}_s\gamma^{\text{Ni}}+m^{\text{Ni}}_s\gamma^{\text{Fe}}}
\label{eq12}
\end{equation}
where $\alpha^{i}$ (i = Ni,Fe) is the pure element damping.
To compare the estimated value of damping $\alpha^{\text{Py}}$ from HHG experiment to the one as a composition of pure elements $\alpha^{\text{Py}}_{\text{eff}}$, we have performed precession measurements in two $10$ nm thick films of pure Ni and pure Fe, using a TMOKE configuration with $25$ fs, $800$ nm pulses. The measured precession damping times $T_d^{\text{Ni, pure}}$, $T_d^{\text{Fe,pure}}$, at fixed initial $20\%$ demagnetization, allows retrieving the corresponding Gilbert damping $\alpha^{\text{Ni,pure}}$ and $\alpha^{\text{Fe,pure}}$ by using equation (\ref{eq12a}). By taking m$_{\text{s}}^{\text{Fe,pure}}$ = 1.72 10$^6$A m$^{-1}$; m$_{\text{s}}^{\text{Ni,pure}}$ = 4.85 10$^6$A m$^{-1}$, the following Gilbert damping values are obtained in pure thin films: $\alpha^{\text{Ni,pure}}$ = 0.05 and $\alpha^{\text{Fe,pure}}$ = 0.016. The  effective damping in Py as a composition of pure elements damping obtained by equation (\ref{eq12}): $\alpha^{\text{Py}}_{\text{eff}}=0.041$ (with $\gamma^{\text{Fe}}$ = 2.12 10$^5$ m s$^{-1}$A$^{-1}$ and $\gamma^{\text{Ni}}$ = 2.03 10$^5$ m s$^{-1}$A$^{-1}$) is in good agreement with values found from equation \ref{eq12a}.
Finally, one can notice that the common Gilbert damping value measured at both Fe and Ni M-edges in Py is close to the highest pure Ni value. It indicates that the dissipation of precession is dominated by Ni sublattice contribution. This could be attributed to the higher spin orbit coupling in Ni compared to Fe, giving higher spin lattice dissipative contribution \cite{Bigot2005,Shokeen2017}.\\

\section{Conclusion}

In this work, magnetization dynamics in Py induced by a 1.5 eV femtosecond pump pulse and probed by HH is investigated with chemical selectivity on Ni and Fe sublattices over a wide temporal scale. The role played by the IEI, in the sublattices damping and precession, has been explored in the intermediate spin temperature range. First, we show that demagnetization dynamics measured at M edge of each Fe and Ni sublattices of permalloy is well reproduced in the LLB framework. The pump fluence dependent dynamics of each sublattice is characterized by double exponential decay with characteristics times $\tau_+$ and $\tau_-$ both relying on elemental susceptibilities and longitudinal damping. This approach allows to distinguish two contributions to the demagnetization time measured at M edges of each sublattice: the first one corresponds to intrasublattice dissipation governed by longitudinal damping and magnetic susceptibilities, the second one is the IEI mediated dissipation responsible of the strongly coupled response observed in this study. An interesting prospective could be to study magnetization dynamics beyond this range of excitation densities, where both sublattices are expected to show different dynamics as predicted by LLB model.

Secondly, we have shown that not only the longitudinal magnetization dynamics of each sublattice is dominated by IEI but also the magnetization vector orientation through precession and transverse damping. The strong IEI drives the two sublattices to share a single precession mode of which the damping is a composition of pure elements damping. Fundamentally, those results improve the understanding of the role of the exchange interaction on ultrafast magnetization dynamics. Moreover, it opens perspectives in the design of new complex magnetic materials for data processing such as alloys and multilayers ferromagnets.
\section*{Acknowledgments}
The authors would like to thank G. Versini for sample preparation, G. Dekyndt, N. Beyer, G. Versini, M. Albrecht and J. Faerber  for technical assistance. This work has been supported by the European Research Council under the project ERC-2009-AdG-20090325 $\sharp$247452 and Agence Nationale de la Recherche (ANR)(ANR-10-EQPX-52).


\end{document}